\documentclass[aps,prd,twocolumn,showpacs,superscriptaddress,groupedaddress]{revtex4}

\usepackage{graphicx}
\usepackage{dcolumn}
\usepackage{amssymb}
\usepackage{amsthm}
\usepackage{amsmath}
\usepackage{bm}
\usepackage{savesym}
\savesymbol{corresponds}
\usepackage{mathabx}
\restoresymbol{ABX}{corresponds}
\usepackage{hyperref}
\usepackage{float}

\begin{document}

\title{Space-time measures for subluminal and superluminal motions}
\author{Benjam\'{\i}n Calvo-Mozo}
\affiliation{Universidad Nacional de Colombia, Observatorio Astron\'omico Nacional, Bogot\'a, Colombia}
\email{bcalvom@unal.edu.co}
\date{\today}

\begin{abstract}

In present work we examine the implications on both, space-time measures and causal structure, of a generalization of the local causality postulate by asserting its validity to all motion regimes, the subluminal and superluminal ones. The new principle implies the existence of a denumerable set of metrical null cone speeds, \{$c_k\}$, where $c_1$ is the speed of light in vacuum, and $c_k/c \simeq \epsilon^{-k+1}$ for $k\geq2$, where $\epsilon^2$ is a tiny dimensionless constant which we introduce to prevent the divergence of the $x,  t$ measures in Lorentz transformations, such that their generalization keeps $c_k$  invariant and as the top speed for every regime of motion. The non divergent factor $\gamma_k$ equals $k\epsilon^{-1}$ at speed $c_k$. We speak then of $k-$timelike and $k-$null intervals and of k-timelike and k-null paths on space-time, and construct a causal structure for each regime. We discuss also the possible transition of a material particle from the subluminal to the first superluminal regime and vice versa, making discrete changes in $v^2/c^2$ around the unit in terms of $\epsilon^2$ at some event, if ponderable matter particles follow k-timelike paths, with $k=1,2$ in this case.

\end{abstract}

\maketitle

\section{Introduction}
The theory of relativity in the restricted sense (special relativity) was developed by \citet{Einstein1905} using two fundamental assumptions, the principle of relativity and the invariance of light speed in vacuum for inertial observers in uniform relative motion. Thereafter, \citet{Minkowski} introduced the notion of space-time and of invariant intervals, as well as light cones (past and future) associated to some event, timelike and spacelike vectors. Almost five decades after the development in full of the general theory of relativity \citep{Einstein1916}, \citet{Hawking-Ellis} rewrite it in a formal way using three postulates for their space-time physics, being the first one that of local causality, in which local light cones are fundamental to define the space-time causal structure. As they have pointed out in their work, tachyons are excluded from this formal description of space-time, for they follow spacelike intervals instead of timelike ones associated to current material particles. Apart from this trouble, apparently there is no other physical impediment for the existence of material particles traveling in space at speeds larger than the speed of light in vacuum. This fact led to many researchers to present their works on superluminal particles or tachyons, for instance, \citet{Feinberg1967}, \citet{Fox1969}, \citet{Camenzind1970}, \citet{Bers1971}, \citet{Asaro1996}, \citet{Recami2000}, \citet{Liberati2002}, \citet{Ehrlich2003}, \citet{Hill-Cox2012}, and many others. In some of these works, Lorentz radical is changed to $\sqrt{v^2/c^2-1}$ in order to keep it a real number. Further, they partition speed values for material particles in two ranges, namely, the subluminal regime $0\leq v<c$ and the superluminal one $c<v<\infty$. This kind of superluminal regime clearly does not have an upper bound. 

Present work proposes that the local causality principle of space-time physics is valid for all motion regimes, the subluminal and superluminal ones. This implies the existence of at least a second metrical null-cone characterized with a speed $c_2>c$ while maintaining the first one, i.e. the light cone with speed $c_1=c$. However, if one allows that superluminal speeds take all positive values, then under present hypothesis there could exist also other metrical null-cones with speeds $c_3$, $c_4$, etc., so that we can say that there are many null-cones, each one with an associated speed $c_k$ with $k\geq1$, a natural number. Einstein's velocity addition rule implies that light speed in vacuum is the top speed in the subluminal regime of speeds; further, light speed in vacuum is the speed associated to the first metrical null-cone. Therefore, if there are other metrical null-cones as we propose here, then their associated speeds, $c_k$ with $k\geq2$, must also follow a rule for velocity addition which keeps $c_k$ as the maximum speed for their respective speeds range, with speed values in the interval $c_{k-1}<v\leq c_k$. We include here  the top speed in these speed ranges, for Einstein's velocity addition rule is valid for speeds in the interval $0\leq v\leq c$, though Lorentz transformations do not include the case $v=c$, which is compatible with the impossibility to find a reference frame at the speed of light in vacuum. Thus, in present work we partition speed values in two kind of ranges, the subluminal one with speeds in the interval $0\leq v\leq c$, a closed interval including both endpoints, and an indefinite large number of superluminal ranges, each one with speeds in the half-closed interval (right-closed) $c_{k-1}<v\leq c_k$, with $k\geq2$.

In next section we deal with the subluminal regime of motions. We begin it by exploring a generalization of Lorentz transformations, which keeps invariant light speed in vacuum and also become finite at this limit speed. We carry out it in two steps. First, we include a very tiny constant (say, $\epsilon^2$) within the square root in the terms where it appears in Lorentz transformations, in order that such terms become finite at the speed of light; we call here $\gamma_1$ to the non-divergent term. As a second step, we change the equations for $y$ and $z$, because we want that the new set of space and time measures let invariant light speed in vacuum. The new expressions for $y$ and $z$ are such that $y/y'$ and $z/z'$ equals $\beta_1\gamma_1$, where $\beta_1$ is the old radical this time in the numerator. We can write down our set of space-time equations in a matrix form, namely, as the factor $\beta_1\gamma_1$ times an inverse Lorentz matrix; therefore, we can say that our set of space-time measures can be described as obtained from a Lorentz transformation followed by a ``regularization" transformation. Then, we found the inverse equations through a matrix process and derive the expressions for moving rods and clocks. Einstein's rule for velocity addition remains the same. At light speed, the factor $\beta_1$ vanishes while $\gamma_1$ equals $\epsilon^{-1}$. The metric is conformal to a Minkowkian one with a conformal factor $\beta_1^2\gamma_1^2$. Our new space-time measures approach Lorentz transformations for most relativistic speeds, i.e. when $\beta_1^2>>\epsilon^2$.

The following section deals with superluminal motions. We begin considering the first superluminal regime, which holds for the speeds interval $c_1< v\leq c_2$, with $c_1=c$; then we develop the appropriate space-time measures for all superluminal regimes. The corresponding space-time equations have the same algebraic form as the generalized Lorentz transformations introduced here for the subluminal regime. For superluminal regimes, we have factors $\beta_k$, $\gamma_k$, $k\geq2$, where in a similar way as we do here with the subluminal regime, we impose that the factor $\beta_k$ vanishes at the speed $c_k$ whilst $\gamma_k=k\epsilon^{-1}$ at that top speed, for all motion regimes. These conditions emerge from the supposition that there is a top speed $c_k$ for the respective interval of speeds. In this way we obtain a generalized dimensionless rule for velocity addition which warrants the existence of that top speed. We give the form of all $\gamma_k$ factors and with the condition of its value at the top speed, we infer that $c_k\simeq\epsilon^{-k+1}c$  for all $k\geq2$. Therefore, we have that $c_{k+1}>>c_k$ for $k\geq1$ because $\epsilon^2$ is a very tiny quantity. We include in this section the values of lengths of moving rods and of time intervals of clocks in motion.

We make a description of the causal structure for all motion regimes in the next section. We define first a k-quadratic form for every pair of events, and later we introduce the here so called ``k-interval", taking into account $c_k^2 dt^2$ and the squared differentials of $x$, $y$, and $z$, adding them with negative sign. Thus, our k-intervals are of types k-timelike or k-null, depending if they are positive or zero, respectively. We define also k-timelike and k-null curves. Given some space-time event, say $p$, we find the region of space-time which can be joined from $p$ with k-timelike paths following the arrow of time, denoted here $I_k^+(p)$, which we call the k-chronological future of $p$, and the future oriented k-null cone associated to it, denoted here $N_k^+(p)$. The union of the two give the set of events $J_k^+(p)$, which can be called the k-causal future of $p$. Similarly, we construct also the sets of events which arrive to $p$ following the arrow of time, through either k-timelike or k-null paths, to obtain the sets $I_k^-(p)$ (the k-chronological past of $p$) and $N_k^-(p)$, respectively, and its union the set $J_k^-(p)$, called here the k-causal past of $p$. Achronal hyper-surfaces are defined for an event $p$ as the set of all events for which the k-quadratic form mentioned above become negative for all positive integers $k$.

Finally, we make an interpretation of the form which acquire our space-time measures for each speed $c_k$, $k\geq1$, associated to every k-null cone. We also estimate a value for our new tiny constant as $\epsilon^2\sim10^{-54}$. In the final discusion we examine the possibility of a discrete transition around the speed of light value for a material particle, such that $v^2/c^2$ could change in terms of $\epsilon^2$ around the unit in both senses, if ponderable matter particles ever follow k-timelike paths; $k=1,2$ in this case.

\section{Subluminal motions}

In present work we assert that the principle of local causality holds at all motion regimes, that is, valid for both, subluminal and superluminal motions. Then, we infer the existence of not only light speed as the top speed for subluminal motions but also the existence of at least a second speed, say $c_2>c$, which serves as the top speed for the first range of superluminal motions. If we accept that superluminal motions consider all speeds greater than light speed, then there should exist also many top speeds for superluminal motions, say $c_3>c_2$, $c_4>c_3$, and so on, in order to preserve local causality. In this way, speeds are partioned in intervals such that the subluminal regime of motions goes in the range of speeds $0\leq v\leq c$, which is a closed interval of real numbers, while superluminal motion regimes go in ranges of speeds which are semiopen intervals of the type $c_{k-1}< v\leq c_k$ for $k\geq2$; in this notation we make $c_1=c$. We construct then space-time measures which for all regimes imply a dimensionless Einstein's rule for velocity addition. We have that this rule includes all speeds $c_k$, $k\geq1$; based on this property, we will also consider space and time measures valid for all speeds in the respective interval of speeds associated to the motions regime, which is given by the positive integer $k$, where $k=1$ denotes the subluminal motions regime.

We start with the subluminal regime of motions. In this regime, we construct space and time measures which preserve the invariance of light speed in vacuum and which do not diverge at the light speed value. The last condition does not imply the existence of relative motions between inertial frames at that top speed, because from any inertial frame light speed takes the same value due to its invariance. The choice of no divergence at light speed for our space-time measures is to assure that both, the space-time measures and the rule for velocity addition holds for the complete interval of speeds of this regime of motions. Further, as we shall see later, it gives us a way to calculate the top speed for every superluminal regime. Let $\epsilon^2$ be a real positive dimensionless constant, very small with respect to the unit; later in this paper we will estimate its value as of the order of $10^{-54}$ -cf. eq.(\ref{epsilon2}). Let us add this constant within the square root which appears in Lorentz transformations, so that our new non-divergent $x$ and $t$ measures are:
\begin{equation}\label{Leps-1}
x=\frac{x'+uct'}{\sqrt{1+\epsilon^2-u^2}},\quad ct=\frac{ct'+ux'}{\sqrt{1+\epsilon^2-u^2}},
\end{equation}
where $u=v/c$ is the dimensionless speed of some inertial frame (primed variables) which is moving uniformly with respect to another inertial frame along the $x$-axis in the positive sense. To preserve light speed in vacuum as the maximum and invariant speed for all inertial frames, we ought to modify the expressions for $y$ and $z$ as well:
\begin{equation}\label{Leps-2}
y=\left(\frac{1-u^2}{1+\epsilon^2-u^2}\right)^{1/2} y',\,\, z=\left(\frac{1-u^2}{1+\epsilon^2-u^2}\right)^{1/2} z'.\\
\end{equation}
We see that eqs.(\ref{Leps-1}),(\ref{Leps-2}) approach Lorentz transformations for most relativistic speeds, that is when $1-u^2>>\epsilon^2$ holds. We can rewrite eqs.(\ref{Leps-1}),(\ref{Leps-2}) in an algebraic form which enables their generalization to other motion regimes. First, let us call $\gamma_1$ to the non divergent term in these expressions:

\begin{equation}\label{gamma1}
\gamma_1 = (1+\epsilon^2-u^2)^{-1/2},\quad u^2=v^2/c^2.
\end{equation}

This gamma factor equals $\epsilon^{-1}$ when $v=c$, which is a large quantity but anyway a finite one. With this gamma factor ($\gamma_1$) our $x$, $t$ measures are:

\begin{equation}\label{ecsxt1}
x = \gamma_k(x'+u_kc_kt'),\quad c_kt =\gamma_k(c_kt'+u_kx'),
\end{equation}

\noindent where $k=1$ for the subluminal regime, for which $c_1=c$, $u_1=v/c_1$, and the $\gamma_1$ factor is given by eq.(\ref{gamma1}). The expressions for the $y$ and $z$ measures in the new algebraic form are:

\begin{equation}\label{ecsyz1}
y = \beta_k\gamma_k y',\quad z = \beta_k\gamma_k z',\quad \beta_k=(1-u_k^2)^{1/2},
\end{equation}

\noindent where $k=1$ for the subluminal regime of motions, that is, for $0\leq v\leq c$. We see that for $v=c$, our set of eqs.(\ref{ecsyz1}) imply the vanishing of $y,z$ measures, while at that speed eqs.(\ref{ecsxt1}) give,

\begin{equation}\label{luz_1}
x = ct= \epsilon^{-1}(x'+ct').
\end{equation}

From eqs.(\ref{ecsxt1}),(\ref{ecsyz1}) we can obtain Einstein's rule of velocity addition in a compact and dimensionless form:

\begin{equation}\label{speeds-sum}
U^{2}_{k}=1-\frac{(1-U'^{2}_{k})(1-u^{2}_{k})}{(1+{\bf u_{k}}\cdot{\bf U'_{k}})^{2}},
\end{equation}

\noindent where $\mathbf{u_k}\cdot\mathbf{U'_k}$ stands for a dot (scalar) product between 3-vectors of dimensionless velocities, ${\mathbf u_k}=\mathbf{v}/c_k$, $\mathbf{U'_k}=\mathbf{V'}/c_k$, and the Cartesian components of vector $\mathbf{V'}$ are $dx'/dt'$, $dy'/dt'$, $dz'/dt'$; the derivatives of the respective unprimed variables leads to the dimensionless speed $U_k$. Vector $\mathbf{v}$, can be considered as usual, that is, as the uniform relative velocity between two inertial observers, except when its magnitude equals the speed of light in vacuum, $c$, which we interpret later.\\

The set of eqs.(\ref{ecsxt1}),(\ref{ecsyz1}) for space and time measures, can be written in a matrix form and from it we can derive their inverses. In effect, if L is the matrix associated to the complete set of these equations, in a matrix form they are $X=LX'$, where $X$,$X'$ are column vectors with $X=(c_kt,x,y,z)^T$, in which $T$ denotes the transpose operation, and $X'$ stands for the respective primed variables column vector. We can easily check that the determinant of the $4\times4$ matrix $L$ equals $\beta_k^4\gamma_k^4$, which does not vanish for $v\neq c_k$; then, we can invert matrix $L$ and obtain the inverse of eqs.(\ref{ecsxt1}),(\ref{ecsyz1}) by means of $X'=L^{-1}X$. In this way, the inverse space-time measures are:

\begin{align}\label{inversas1}
c_kt'= &\,\beta_k^{-2}\gamma_k^{-1}(c_kt-u_kx),\\ 
x'= &\,\beta_k^{-2}\gamma_k^{-1}(x-u_kc_kt),\label{inversas1b}\\ 
y'= &\,\beta_k^{-1}\gamma_k^{-1}y,\,\, z'=\beta_k^{-1}\gamma_k^{-1}z.\label{inversas2}
\end{align}

Let us observe that in the subluminal motions regime these expressions reduce to the well known ones associated to Lorentz transformations for $\beta_1^2>>\epsilon^2$. In the previous notation, Lorentz transformations are usually written in a matrix form as $X'=\Lambda X$, so that $X=\Lambda' X'$, where $\Lambda'$ is the inverse matrix of $\Lambda$.  Let us write now our set of eqs.(\ref{ecsxt1}),(\ref{ecsyz1}) in a matrix form as:
\begin{equation}\label{Lk}
X=\beta_k\gamma_k\Lambda_k'X'=L_k X', \quad L_k=\beta_k\gamma_k\Lambda_k',
\end{equation}
where $\Lambda_k'$ is a generalization of the Lorentz matrix and equals to it for $k=1$. For any integer $k\geq1$ one has:
\begin{equation}\label{Lmatrix}
\Lambda_k'=
\begin{bmatrix}
\beta_k^{-1} & u_k\beta_k^{-1} & 0 & 0 \\
u_k\beta_k^{-1} & \beta_k^{-1} & 0 & 0 \\
0 & 0 & 1 & 0 \\
0 & 0 & 0 & 1
\end{bmatrix}
\end{equation}
The inverse matrix of $\Lambda_k'$, $\Lambda_k$, is obtained from eq.(\ref{Lmatrix}) changing the sign of $u_k$. Thus, the inverse eqs.(\ref{inversas1})-(\ref{inversas2}) can also be expressed as $X'=\beta_k^{-1}\gamma_k^{-1}\Lambda_k X$, $u_k\neq1$. From these equations we can obtain a rule to add speeds, which in a compact form is similar to eq.(\ref{speeds-sum}), interchanging $U_k$ and $U'_k$, and changing the term with the dot product there by $-{\bf u_{k}}\cdot{\bf U_{k}}$. We can also write down eqs.(\ref{ecsxt1}), (\ref{ecsyz1}) in a vector form as:
\begin{equation}
\begin{gathered}\label{vector-form}
{\bf r}= \beta_k\gamma_k{\bf r}' + \gamma_k\Bigl[(1-\beta_k)\frac{{\bf u}_k\cdot {\bf r}'}{u_k^2}+c_k t' \Bigr]{\bf u}_k,\\
c_k t= \gamma_k\left( c_k t' +{\bf u}_k\cdot{\bf r}' \right).
\end{gathered}
\end{equation}
Then, for $v=c_k$ there is only one degree of freedom:
\begin{equation}\label{vector-c_k}
{\bf r}=c_k t\, \hat{{\bf e}}=k\epsilon^{-1}\left({\bf r}'\cdot\hat{{\bf e}} + c_k t'\right)\hat{{\bf e}},
\end{equation}

\noindent where $\hat{{\bf e}}={\bf u}_k/u_k$ denotes the direction of propagation.

For subluminal motions, we can calculate lengths of moving bodies in the customary relativistic way, that is, measuring simultaneously their extreme points; thus, with the aid of eqs.(\ref{inversas1b}),(\ref{inversas2}) we obtain for lengths parallel to the motion and perpendicular to it, the values $l_{\parallel}=\beta_1^2\gamma_1 l_o$ and $l_{\perp}=\beta_1\gamma_1 l_o$, respectively, where $l_o$ is the rest length; then, volume of moving bodies varies as $\beta_1^4\gamma_1^3$ times the volume at rest. For clocks in motion, time intervals are seen as $\gamma_1\Delta t_o$, where $\Delta t_o$ is the corresponding lapse of proper time. However, all expressions considered here reduce to the respective relativistic ones when $\beta_1^2>>\epsilon^2$. The appearance of the regularization factor $\beta_1\gamma_1$ in our space-time measures, see eqs.(\ref{Lk}), lead us to consider the existence of a preferred inertial frame. For instance, we can think of either a frame associated to the cosmic microwave background corrected by the dipole  \cite{CMB_dipole},\cite{CMB_spec} or a ``cosmic rest frame" of comoving observers in a FLRW model \cite{cosmic_restframe}.

\section{Superluminal motions}

As we have mentioned above, in present work the author proposes that besides the causal structure of space-time for subluminal motions, there could be another causal structure for superluminal motions. This hypothesis implies the existence of at least a speed very much larger than the speed of light in vacuum, say, $c_2>>c$, which for inertial observers in uniform relative superluminal motion behaves as an invariant and top speed, which in turns can be associated to a {\it new metrical null cone}. In effect, this new null cone enables one to think of superluminal timelike intervals and curves as we deal with in the next section. But if as many researchers working on superluminal motions think (see, e.g. \citet{Hill-Cox2012}), material particles in superluminal motion or tachyons have speeds in the open interval of reals $1<v/c<\infty$, then one could also admit the possible existence of other invariant and top speeds, say, $c_3>>c_2$, $c_4>>c_3$, etc., in such a manner that speeds are partitioned in intervals of the type $0\leq v\leq c$ for subluminal motions, and $c<v\leq c_2$, $c_2<v\leq c_3$, and so on, for superluminal motions. We have included the top speeds for each interval of speeds or ``regime of motions" for two reasons, namely, because (i) these top speeds are admitted by the generalized dimensionless rule for velocity addition, and (ii) for if we accept them as special cases in our space-time measures, making these measures non-divergent at $v=c_k$, then it results a rule for finding the values of each top speed $c_k$ for every $k\geq2$.

Let us consider initially the first range of superluminal motions; the set of space-time measures for this motions regime should imply a similar (in algebraic form) composition rule for dimensionless speeds but which takes into account$\ c_2$ instead of $c$. We know that eqs.(\ref{ecsxt1}),(\ref{ecsyz1}) infer Einstein's rule for velocity addition given here by eq.(\ref{speeds-sum}) in a compact form. Thus, we need corresponding $\beta_2$ and $\gamma_2$ factors, where the subindex $k=2$ stands for superluminal motions under the $c_2$-cone regime. For the $\beta_2$ expression we have that it vanishes when $v=c_2$ as we infer from the third of eqs.(\ref{ecsyz1}). To find an expression for $\gamma_2$, we take into account that $\gamma_1=\epsilon^{-1}$ at $v=c$, and that the new gamma factor, $\gamma_2$, increases as $u^2$ goes up, till one reaches the new maximum speed, say, $c_2$. As the gamma factor $\gamma_1=\epsilon^{-1}$ at the speed $c_1=c$, we reasonably assume that $\gamma_2=2\epsilon^{-1}$ when $v=c_2$, and in general, $\gamma_k=k\epsilon^{-1}$ at $v=c_k$. We can obtain this result (for $k=2$) combining positive and negative powers of $\epsilon^2$:

\begin{equation}\label{gamma2}
\gamma_2 =\epsilon^{-1}+(\epsilon^{-2}+1+\epsilon^2-u^2)^{-1/2}.
\end{equation}

From this expression we obtain the new limit speed for the first superluminal regime, $c_2$, by considering that $\gamma_2$ equals $2\epsilon^{-1}$ at $v=c_2$:

\begin{equation}\label{c_2}
c_2=(\epsilon^{-2}+1)^{1/2}c\simeq\epsilon^{-1}c \sim5\times10^{26}c.
\end{equation}

This is a huge speed. The value of $v=c_2$ given above takes into account the estimative value of $\epsilon^{-2}$ as given by eq.(\ref{epsilon2}). In our description, the $\gamma_1$ factor given by eq.(\ref{gamma1}) is valid for squared dimensionless speeds in the range $0\leq u^2\leq 1$, that is, in the subluminal regime, while the $\gamma_2\,$factor given by eq.(\ref{gamma2}) applies in the range of squared dimensionless superluminal speeds $1< u^2\leq\epsilon^{-2}+1$; in both cases $u=v/c$. Lengths of moving bodies and time intervals given by clocks in this superluminal regime can be derived using the same procedures as done for the subluminal case: $L_{\parallel}=\beta_2^2 \gamma_2 l_o$, $L_{\perp}=\beta_2\gamma_2l_o$, $\Delta t=\gamma_2\Delta t_o$. These expressions have an interesting behaviour as, for in all of them appears a factor on the order of $\epsilon^{-1}$ for most conceivable speeds. In effect, if $u_2^2\ll1$, then $\beta_2\simeq(1-u_2^2/2)$ and $\gamma_2\simeq(\epsilon^{-1}+\epsilon/\beta_2)$, which in turns can be approximated to the unit and to $\epsilon^{-1}$, respectively. At the second limit speed, $c_2$, eqs.(\ref{ecsxt1}),(\ref{ecsyz1}), give $x=c_2t=2\epsilon^{-1}(x'+c_2t')$, which is similar in form to eq.(\ref{luz_1}) except for a factor of 2, whilst measures $y,z$ vanish.

We will find now the appropriate expressions for $\gamma_k$, $c_k$; the one for $\beta_k$ is given by eq.(\ref{ecsyz1}). We see that eqs.(\ref{ecsxt1}),(\ref{ecsyz1}) imply eq.(\ref{speeds-sum}) for speed addition in a compact form, which let $c_k$ as the maximum speed in the interval $c_{k-1}<v\leq c_k$, for any integer $k\geq2$. We obtain $\gamma_k$ for any $k\geq2$, using only positive and negative powers of $\epsilon^2$, considering that it has an accumulated $(k-1)\epsilon^{-1}$ from the previous speeds range and that it should equal to $k\epsilon^{-1}$ at the top speed $c_k$ of the respective speeds interval. Further, as in the expression for $\gamma_2$ there are terms with positive and negative powers of $\epsilon^2$, then we shall use higher positive and negative powers of it keeping ``symmetry" in these powers, that is, if there appears the power $\epsilon^{-2j}$ then there appears also the power $\epsilon^{2j}$. Using these considerations we have ($k\geq2$):

\begin{equation}\label{gammai}
\gamma_k =(k-1)\epsilon^{-1}+ \epsilon^{k-2}\left[\sum_{j=1}^{k-1}\left(\epsilon^{-2j}+\epsilon^{2j}\right)+1-u^2\right]^{-1/2}.
\end{equation}
\\
The condition imposed on the factor $\gamma_k$ at the maximum speed of the associated range, gives us the expression of $c_k$ for $k\geq2$:

\begin{equation}\label{luz-i}
(c_k/c)^2 =\sum_{j=1}^{k-1}\epsilon^{-2j} + \sum_{j=0}^{k-2}\epsilon^{2j},
\end{equation}

\noindent which reduces to the expression given by eq.(\ref{c_2}) for $k=2$. For any $k\geq2$ eqs.(\ref{c_2}),(\ref{luz-i}) tell us that in first approximation $c_k\simeq\epsilon^{-k+1}c$.  For all $c_k$ one has that $y=z=0$, and:

\begin{equation}\label{xt-luz-i}
x=c_kt=k\epsilon^{-1}(x'+c_kt'),\quad \text{and} \quad x'=c_k t',
\end{equation}

\noindent which is a generalization of eq.(\ref{luz_1}). The first of eqs.(\ref{xt-luz-i}) is obtained directly from eqs.(\ref{ecsxt1}) making $u_k=1$, that is, for $v=c_k$. For that value eqs.(\ref{ecsyz1}) give $y=z=0$. The second of eqs.(\ref{xt-luz-i}) is obtained through a limit procedure, making use of the l'H\^opital rule of calculus applied to eqs.(\ref{inversas1}),(\ref{inversas1b}).\\

\section{Causal structure}
Let $\mathcal{M}$ be the set of all events in space-time; every point of it has four components, such that each point or event $p$ can be written in components as $(x^0,x^1,x^2,x^3)$, of which the former contains time, $x^0=c t$. We can define on $\mathcal{M}$ a quadratic form generalizing the one proposed by \citet{Zeeman1964}, and \citet{Kronheimer-Penrose1967}, such that if $p$ and $q$ are events, with components $x^{\alpha}$, $y^{\beta}$, respectively, with $\alpha,\beta:0,1,2,3$, then:
\begin{equation}\label{2-form}
Q_k(p,q)= (c_k/c)^2(x^0 - y^0)^2 - \sum_{i=1}^{3}(x^i - y^i)^2.
\end{equation}

If the event $q$ lies in some small open neighborhood of the event $p$, there exists the possibility to have a differential which can be called a ``k-interval":
\begin{equation}\label{k-interval}
ds^2_k=c^2_k dt^2 - dx^2 - dy^2 - dz^2,
\end{equation}
\noindent where $c_k dt$, $dx$, $dy$, and $dz$ can be obtained from eqs.(\ref{ecsxt1}),(\ref{ecsyz1}) with $u_k$, $\beta_k$, $\gamma_k$ constants; $c_k$ is given by eq.(\ref{luz-i}) for $k\geq2$ and $c_1=c$. For each $k\geq1$, $k$ an integer, one has a motions regime, that is, an interval of speeds for which there is a top and invariant speed $c_k$ and a set of space-time measures given by eqs.(\ref{ecsxt1}),(\ref{ecsyz1}). The k-intervals are said to be k-timelike or k-null depending if the value of $ds^2_k$ as given by eq.(\ref{k-interval}) is positive or zero, respectively. From eq.(\ref{k-interval}) one obtains a metric ${\bf g}_k$ for every motions regime, which results to be conformal to a Minkowskian metric $\mbox{\boldmath $\eta$}$ for the primed variables:
\begin{equation}\label{k-metric}
{\bf g}_k=\beta_k^2\gamma_k^2\mbox{\boldmath $\eta$}.
\end{equation}

The notions of causal precedence ($\prec$) and of chronological ($\lll$) precedence as developed by \citet{Kronheimer-Penrose1967}, can be generalized for all motion regimes distinguishing them with an index $k\geq1$, taking into account the causality and chronological relations as described by \citet{Carter1971}, such that given two events of the space-time manifold, say, $p,q\in \mathcal{M}$, then $p$ causally precedes $q$ with respect to some ``auxiliary set" $\mathcal{U}\subset \mathcal{M}$,

\begin{equation}\label{causal-k}
p\underset{\mathcal{U}}{\prec} q,\, \text{if}:\, x^0<y^0\,\, \text{and}\,\, Q_k(p,q)\geq0,
\end{equation}

\noindent under the condition to be restricted to the subset $\mathcal{U}\subset\mathcal{M}$. For subluminal motions, $k=1$, and our set $\mathcal{U}$ is confined to the light cone and its interior; let us denote it as $\mathcal{U}_1$. For $k=2$, the first superluminal regime, the auxiliary set $\mathcal{U}$, say, $\mathcal{U}_2$, is restricted to the region within the light cone (excluding it) and the second null cone, including it. In general, for any motion regime distinguished by some $k\geq2$, the corresponding set $\mathcal{U}_k$ is restricted to the region within the ($k-1$)-null cone, excluding it, and the $k$-null cone taking it as part of this reference set. Then, taking into account all motion regimes, labelled by an integer $k\geq1$ we have the following partial orderings in space-time for all cases:

\begin{equation}\label{k-relations}
\begin{split}
p\underset{\mathcal{U}_k}{\prec} q, &\quad \text{if}:\, x^0\leq y^0\,\, \text{and}\,\, Q_k(p,q)\geq0,\\
p\underset{\mathcal{U}_k}{\lll} q, &\quad \text{if}:\, x^0< y^0\,\, \text{and}\,\, Q_k(p,q)>0,\\
p\underset{\mathcal{U}_k}{\to} q, &\quad \text{if}:\, x^0\leq y^0\,\, \text{and}\,\, Q_k(p,q)=0.
\end{split}
\end{equation}

In analogy with the definitions given by \citet{Kronheimer-Penrose1967}, we can call the above relations, k-causal precedence, k-chronological precedence and k-horismos, respectively. In the respective subset of space-time associated to some pair of events in the same motion regime, in which one of them causally or chronologically precedes the other, one can link the two through a continuous succession of events, that is, we link them by a curve in space-time. Any curve on the space-time manifold is conceived as usual, that is, as a map of an interval of the reals on the space-time manifold; if the curve does not intersect with itself, then it is a simple curve. In present work we say that in space-time a curve is a k-timelike curve if every pair of events on it, say, $p,q\in\lambda_k$, where $\lambda_k$ is the k-timelike curve, are such that $Q_k(p,q)>0$. We say also that a $\mu_k$ curve is a k-null curve if for every pair of events on it, $p,q\in\mu_k$, one has that $Q_k(p,q)=0$. If these curves are future directed, one has further that if they go from event $p$ to event $q$, then $x^0<y^0$.\\

\begin{figure}
\centering
\includegraphics[scale=0.4]{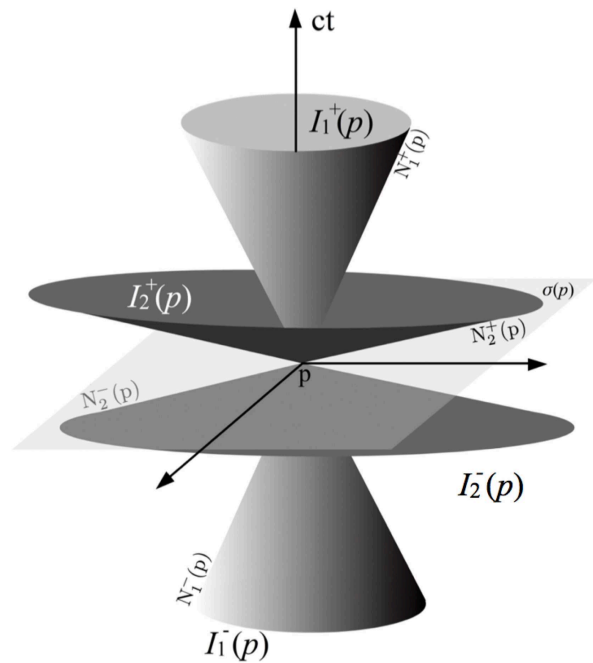}
\caption{Given some event $p$ one has their associated past and future $c_1$ and $c_2$-cones, and the corresponding 1-,2-chronological past and future regions.}\label{null-cones}
\end{figure}

Under the previous partial orderings in space-time and given a motion regime labeled by $k$, we see that if the event $p$ chronologically precedes the event $q$, then there is at least one k-timelike curve joining them from $p$ to $q$ as time runs forwardly. Similarly, if the event $p$ causally precedes the event $q$, then as time goes on, there is either a k-timelike curve or a k-null curve in the sense from $p$ to $q$. Thus, for any event $p\in\mathcal{M}$, we define the following regions of space-time which lead us to the notion of the causal structure of space-time for each regime of motion:
\begin{equation}\label{k-regions}
\begin{split}
I_k^+(p)=& \{q\in\mathcal{M}: p\underset{\mathcal{U}_k}{\lll} q\},\\
I_k^-(p)=& \{q\in\mathcal{M}: q\underset{\mathcal{U}_k}{\lll} p\},\\
J_k^+(p)=& \{q\in\mathcal{M}: p\underset{\mathcal{U}_k}{\prec} q\},\\
J_k^-(p)=& \{q\in\mathcal{M}: q\underset{\mathcal{U}_k}{\prec} p\},\\
N_k^+(p)=& \{q\in\mathcal{M}: p\underset{\mathcal{U}_k}{\to} q\},\\
N_k^-(p)=& \{q\in\mathcal{M}: q\underset{\mathcal{U}_k}{\to} p\},\\
\end{split}
\end{equation}
which can be called the k-chronological future, k-chronological past, k-causal future and k-causal past of the event $p$, respectively, for the first four subsets of space-time associated to $p$, and for the last two above we call them the future and past k-null cones, respectively. In Fig.\ref{null-cones} we represent the first two past and future metrical null cones associated to some event $p\in\mathcal{M}$; they are not drawn to scale and are labeled as the $N_1^+(p)$, $N_2^+(p)$ regions in the upper part of the figure and the regions $N_1^-(p)$, $N_2^-(p)$ in the lower part of it. We also see there the regions of k-chronological past and future of the event $p$, with $k=1,2$, that is, for the subluminal ($I_1^-(p)$, $I_1^+(p)$) and first superluminal ($I_2^-(p)$, $I_2^+(p)$) motion regimes.

According to \citet{Kronheimer-Penrose1967} the set $\mathcal{M}$ together with the partial orderings like our k-causal precedence, k-chronological precedence and k-horismos, constitute a causal space; in our case, causality in space-time is by regions, each one corresponding to a motion regime denoted by a natural number $k$; in this way, our space-time is causally multi-structured. There are also events which belong to the achronal hypersurface of $p$, that is, events which are outside any k-causal past or future region of $p$. Thus, we define the achronal hypersurface associate to the event $p$ by:
$$
\sigma(p)= \{ q\in\mathcal{M}: Q_k(p,q)<0,\,\, \forall k\in\mathbb{N}\},
$$

\noindent where $\mathbb{N}$ is the set of natural numbers. In Fig.\ref{null-cones} we show also the hypersurface $\sigma(p)$. If events $p',p''\in\mathcal{M}$ belong to $\sigma(p)$, then their achronal hypersurfaces are congruent with $\sigma(p)$. Therefore, we can define achronal hypersurfaces independently of some given event as:
\begin{equation}\label{achronal-s}
\sigma= \{ p\in\mathcal{M}:\exists q\in\mathcal{M},\, Q_k(p,q)<0,\,\, \forall k\in\mathbb{N}\}.
\end{equation}

These spacelike hypersurfaces are 3-dimensional manifolds or {\it slices} of space-time. Let $\sigma_1,\sigma_2$ be two achronal hypersurfaces. Then, if the points (i.e. events) in one of them are reachable from the other through future directed k-timelike or k-null paths, we can label each achronal surface $\sigma$ by some time $t(\sigma)$, based on Geroch's splitting theorem \cite{Geroch1970} applied to our flat space-time, such that $t(\sigma_2)>t(\sigma_1)$ if points on $\sigma_2$ are reachable from some point on $\sigma_1$ through future directed k-null or k-timelike paths. To see it, we can construct as many k-null cones as desire at every point on $\sigma_1$ and look their intersection with the other hypersurface $\sigma_2$. Intuitively, we can take a look at Fig.\ref{null-cones} for a flat 3D space-time (2+1) and imagine that the plane containing the event $p$ there is $\sigma_1$; thus, $\sigma_2$ will be in the upper part or the lower part of that figure. If we choose the first option, then the regions of $\sigma_2$ which belong to either $J_1^+(p)$ or $J_2^+(p)$ are such that their points can be joined from $p\in\sigma_1$ either through future directed k-null or k-timelike paths, with $k=1,2$.

\section{An interpretation}
The appearance of a set of top and invariant speeds by pieces, partition speed values in sets of nonnegative real numbers or intervals which we can call speed ranges, being the first one closed at both sides, that pertaining to the subluminal regime, whilst the others are open in the lower end and closed in the upper value and which correspond to superluminal regimes. This fact and taking also into account that measuring factors $\gamma_k$, $\beta_k$ take the maximum and minimum value, respectively, at the top speed of the respective speeds range, enable us to infer a new property of matter, and is the existence of an indefinite large number of metrical k-null states of matter which determine the behaviour of any material body depending on its speed with respect to those of the k-null states. Thus, we introduce in present work the following postulate:\\

{\em Space-time is locally causal at all motion regimes, which are determined by a denumerable set of speeds associated to metrical null cones},\\

\noindent where a denumerable or countably infinite set is one which has the same cardinality than the set of natural numbers. Thus, if $C=\{c_k: k\in\mathbb{N}\}$ is the set of all speeds associated to metrical null-cones, then we see that for every natural number $k$ there is one speed $c_k$ associated to the respective k-null cone. This postulate is a generalization of the corresponding postulate proposed by \citet{Hawking-Ellis}. It enables one to introduce the quadratic form given by eq.(\ref{2-form}), or more general, the k-interval of eq.(\ref{k-interval}). Lorentz like matrices of the type given by eq.(\ref{Lmatrix}) are necessary for they imply the invariance of the top speed $c_k$ for each motion regime, i.e. for each $k\geq1$. Then, after a Lorentz like transformation space-time measures are subject to a regularization transformation as stipulated by eq.(\ref{Lk}), because the top speed $c_k$ is included in the respective speeds interval in the partition of speeds. Though our eqs.(\ref{ecsxt1}),(\ref{ecsyz1}) do not diverge for $v=c_k$, it does not mean that it serves as a reference frame to describe events, because eq.(\ref{speeds-sum}) derived from these equations imply that anyway it is seen with speed $c_k$ by any inertial observer. These special states of space-time measures have only one degree of freedom, for in that case $y,z=0$, and $x=c_kt$. Further, we have $x'=c_kt'$. To see it, let us examine the light case. If we take the limit $u\to1$ in eqs.(\ref{inversas1}),(\ref{inversas1b}), consider eq.(\ref{gamma1}) for $\gamma_1$, and the third of eqs.(\ref{ecsyz1}) for $\beta_1$, one obtains $x'=ct'=\epsilon x/2$. This result enables us to examine the behaviour of our space-time eqs.(\ref{ecsxt1}),(\ref{ecsyz1}) for light, that is, for eq.(\ref{luz_1}). It tell us that half of the {\it structure} associated to photons lies in $x'$ whilst the other half corresponds to $ct'$. A possible interpretation of this result is that physical measures associated to light reside in every point of that part of it, denoted by $x'$ and measured as $\epsilon^{-1}x'=x/2$; the other half, that pertaining to $ct'$, indicates a tendency to move, that is, to occupy an equal amount of space, just contiguous to the first one, but to be ocuppied a time $\epsilon^{-1}t'=x/(2c)$ later. As both terms appear in eq.(\ref{luz_1}) they are integral part of light structure. Once the former half occupies the second one, it continues this tendency successively,  that is, periodically, and maybe this is the reason of its wavelike behavior. In effect, from the viewpoint of an inertial observer who sees the propagation of some light ray, describing its position by the vector {\bf r} when his clock indicates the time $t$, the physical situation of the propagation of the light ray repeats every time the difference ${\bf r}\cdot\hat{{\bf e}}-ct$ equals $\lambda$, where $\hat{{\bf e}}$ indicates the direction of propagation and $\lambda$ equals the spatial period. If the situation is one in which there is a continuous source of light in the given direction and with spatial period $\lambda$ (light wavelength), then as being a periodic phenomenon, it can be described by means of a periodic function like sine or cosine of an argument or phase containing that difference over $\lambda$, which varies from $0$ to $1$ and then one multiplies it by $2\pi$, for it is the period of basic trigonometric functions. We can redefine $(2\pi/ \lambda)\hat{{\bf e}}$ as the wave vector ${\bf k}$ and calling frequency to $\nu=c/\lambda$, $\omega=2\pi\nu$, and obtain the usual description of monochromatic light as:
$$
{\bf A}\exp \left(i({\bf k}\cdot{\bf r}-\omega t)\right),
$$

\noindent where ${\bf A}$ is the amplitude. For higher null-cones states propagating at speeds $c_k$, $k\geq2$ an integer, we have a similar situation as can be inferred from eqs.(\ref{vector-c_k}),(\ref{xt-luz-i}); therefore, if they exist, their propagation should exhibit a wavelike behaviour, too.

Going forward in our interpretation of the results presented here, we can think that when photons propagate, or particles move, what are moving are their associated structures defined on the elementary bases of space, understanding by them the finest partition of space. When we go down looking for the finest partition of space, we find lengths on the order of Planck's length, $L_P=(\hbar G/c^3)^{1/2}$. \citet{Planck1899} thought that when we arrive to such lengths, space could be described in a discrete manner, while for \citet{Sakharov1967} lengths of order $L_P$ represent our limits of the concept of space in the sense of localization. Wheeler thought of some kind of pregeometry at such levels (see box 44.5 of \cite{Gravitation}) as the ``basic building" of spacetime. For \citet{Smolin2001} our finest partition of space is a very tiny volume given by $L_P^3$ and the finest time interval is $t_P=L_P/c$, the Planck time. Taking into account these ideas, the author of present paper proposes here that what we interpret as a point in our space-time measures given by eqs.(\ref{ecsxt1}),(\ref{ecsyz1}), or of their inverses given by eqs.(\ref{inversas1})-(\ref{inversas2}), is something of the order of $L_P^3$ in volume, which we will call here an {\it element} of space or of the structures under discussion, like those associated to photons or particles, or even of some region of space wherein there exists some field. On such elements of structures associated to photons or particles, or of regions of space with fields, we can define geometrical, kinematical or dynamical measures.

\section{Value of $\epsilon^2$}
To estimate a value for $\epsilon^2$ we see first that when $u=0$, eqs.(\ref{Leps-1}),(\ref{Leps-2}) give us spacetime measures of type $\delta Y/Y'\simeq-\epsilon^2/2$ where $\delta Y=Y-Y'$; in these couple of expressions $Y$ denotes either $x$, $y$, $z$, $ct$, and $Y'$ stands for either of the corresponding primed variables. A possible interpretation is that in any vacuum, even if it is far from the influence of ponderable matter or fields, vacuum has natural fluctuations and then if one put there a test particle it should be influenced by these fluctuations. For instance, our metric is such that $\delta g/g\simeq-\epsilon^2$ in first approximation for $u^2<<1$. In effect, in the strict subluminal ($u<1$) relativistic ($\beta_1^2>>\epsilon^2$) case, our k-metric (cf. eq.(\ref{k-metric})) for $k=1$, approximates to:
\begin{equation}\label{g1-approx}
{\bf g}_1=\left(1+\frac{\epsilon^2}{1-u^2}\right)^{-1}\mbox{\boldmath $\eta$}\simeq\left(1-\frac{\epsilon^2}{1-u^2}\right)\mbox{\boldmath $\eta$}.
\end{equation}

However, we should look for an observable quantity in order to estimate a value for $\epsilon^2$ by analogy. In author's opinion it is the case of the correction made on the electron spin g-factor, $(g_e-2)/2$, associated to the anomalous magnetic moment of electron, first calculated by \citet{Schwinger} and which is on the order of $\alpha/(2\pi)$, where $\alpha$ is fine's structure constant, an effect basically due to radiative corrections from vacuum fluctuations. \citet{Weinberg} employs the concept of ``charge radius" in these calculations as the zone of net influence of vacuum fluctuations on the electron; we use here the Compton wavelength of electron, $\lambda_e\simeq2.4\times10^{-12}\,\text{m}$, as an approximation to the radius of the zone of influence of vacuum fluctuations on the electron, as we infer from \citet{Weisskopf1949}. As pointed out by \citet{Gravitation} at scales of distances comparable to Planck's length $L_P$, vacuum fluctuates, such that if $\delta g$ denotes the fluctuations in metric coefficients, they state that it is of order of $\delta g\sim L_P / l$, where $l$ is a linear scale of distance of the ``region under study". We can make instead, a re-interpretation of Misner, Thorne  and Wheeler previous idea by saying that $\lambda_e/L_P$ gives an integer number of elements, say, $N_o$, as the number of associated vacuum elements which should be taken into account for the observed action of vacuum fluctuations on an electron, for instance affecting its spin magnetic moment. We see that the ratio of electron's Compton wavelength over Planck's length gives on the order of $10^{23}$. In order to put well known numbers, we can take $N_o$ of the order of a familiar quantity, say, Avogadro's number, but taking it dimensionless: $N_o=N_A\times(1\text{mol})$, where $N_A$ is Avogadro's number. According to our interpretation for light (see previous section), the right hand side of eq.(\ref{luz_1}) can be seen as light's wavelength. Thus, and in agreement with de Broglie's hypothesis ($\lambda_{dB}=h/p$), we can expect that the term $\epsilon^2$ from vacuum fluctuations at every vacuum element, multiplies the term $\epsilon^{-1}$ of particle's wavelength, such that the net effect of $N_o$ vacuum elements on the electron, $N_o\epsilon$, gives the observed anomalous correction to electron's spin magnetic moment, i.e. $\alpha/(2\pi)$. That is, $N_o\epsilon=\alpha/(2\pi)$, or:
\begin{equation}\label{epsilon2}
\epsilon^2 = \left(N_o^{-1}\frac{\alpha}{2\pi}\right)^2 \simeq 3.7\times10^{-54},
\end{equation}
\noindent which is an extremely tiny constant. From it, we see that $\epsilon^{-1}\simeq5\times10^{26}$. Then, with these figures in mind and eqs.(\ref{c_2}),(\ref{luz-i}) one easily checks that $c_{k+1}/c_k>>1$ for every $k\geq1$.

\section{Discussion}
It is well known that ponderable matter particles move in space-time along timelike paths. This description is usually done in the subluminal regime of motions. In present work we extended the notion of timelike curves, such that we have k-timelike curves, depending on the regime of motions in which we describe material particles.  Thus, the usual timelike curve is now a 1-timelike curve. Let us consider now the theoretical possibility of a change of the regime of motion for material particles. Clearly, such a hypothetical change can not be done in a continuous manner. If we consider a material particle which has been accelerated from the rest till some speed near but lesser than the speed of light in vacuum, all this process is described in space-time by a 1-timelike path, and if it surpasses the light speed it should be described by a 2-timelike curve; in no moment it moves with light speed during this transition, because if it does, ceases to be a ponderable matter particle. Then, the transition from the subluminal regime to the first superluminal regime of motions around light speed in vacuum, that is, around the $u^2=1$ value, should be done in a discrete way; as the $\gamma_1$,$\gamma_2$ factors contain the constant $\epsilon^2$ as the finest quantity there, then one can think that changes in $u^2$ are of the type:
\begin{equation}\label{cambio_u2}
u^2: 1-n_1\epsilon^2\rightleftarrows1+n_2\epsilon^2,
\end{equation}
\noindent where $n_1$,$n_2$ are positive real numbers, and the cited transition is given by the left to right arrow above. The other direction, the right to left transition, corresponds to a change from the first superluminal regime of motions to the subluminal one. Measurement factors $\gamma_1$, $\gamma_2$ given by eqs.(\ref{gamma1}),(\ref{gamma2}), respectively, enables one to think that the real numbers $n_1$,$n_2$ in eq.(\ref{cambio_u2}) are integers, because before the unit, $u^2=1-\epsilon^2$ in $\gamma_1$. If we have now in this regime some $u^2=1-n_1\epsilon^2$, $n_1$ a natural number, it gives $\gamma_1=(n_1+1)^{-1/2}\epsilon^{-1}$; for instance, if $n_1=3$ one obtains $\epsilon^{-1}/2$. For the first superluminal regime, we can consider that the minimum squared dimensionless speed equals $u^2=1+\epsilon^2$, which corresponds to $\gamma_2=\epsilon^{-1}+\epsilon$. According to  eq.(\ref{cambio_u2}) net changes in $u^2$ are $\pm(n_1+n_2)\epsilon^2$ where the plus sign applies for the subluminal to superluminal transition.
\begin{figure}
\centering
\includegraphics[scale=0.4]{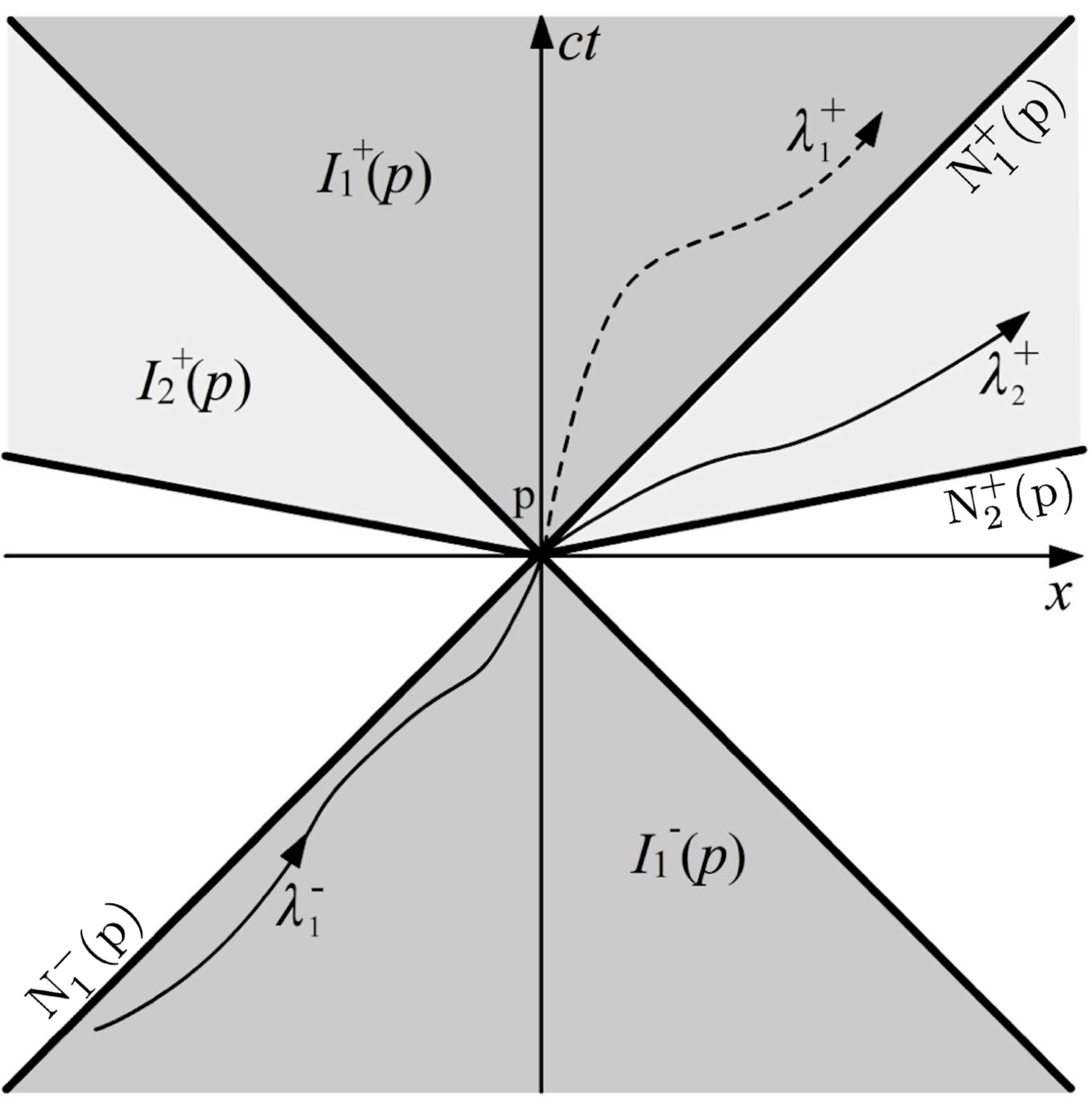}
\caption{A ponderable matter particle could make a transition from the subluminal regime of motions to the superluminal one at some event $p$ through a discrete jump in $u^2=v^2/c^2$ around the unit in terms of $\epsilon^2$.}\label{transition1-2}
\end{figure}\\

We illustrate this transition in Fig.\ref{transition1-2}; there, $p$ is some event of space-time where a transition happens as described by eq.(\ref{cambio_u2}).

One of the goals of NASA's Breakthrough Propulsion Physics Program (BPPP), as reported by \citet{Millis}, was looking for a possible transition around the speed of light value for a material body. The answer was negative, for present day physics does not allow it due to the continuous character of speed changes, which implies an infinite energy consumption. Present work enables to answer that BPPP's question affirmatively, as inferred from our considerations above and eq.(\ref{cambio_u2}). However, our proposition differs from that made by \citet{Alcubierre}. In effect, a ponderable matter particle can start at rest, accelerates to a speed near but lesser than light speed in vacuum, following a 1-timelike curve, and at some event (say, $p$) make a discrete transition in $u^2$ according to eq.(\ref{cambio_u2}), then it changes its path in space-time as a 2-timelike curve. Thus, the incoming 1-timelike curve of the particle, say, $\lambda_1^{-}$, belongs to the region $I_1^-(p)$ and the path after the discrete change in $u^2$, say, $\lambda_2^{+}$ lies in the region $I_2^+(p)$. In Fig.\ref{transition1-2} we represent this conceptual change; the dotted path in the region $I_1^+(p)$ would be the path followed by the particle if no discrete change in $u^2$ occurs at event $p$. Mathematically, $\lambda_1^{-}$ and $\lambda_2^{+}$ coincide at event $p$ but their slopes differ there due to the change described by eq.(\ref{cambio_u2}). This change is done at some ``instant of time", which according to \citet{Smolin2001} is a lapse of time equals to Planck time $t_P=L_P/c$, as it corresponds to the finest partition of time.

\section{Conclusions}
In present article we obtain for the subluminal regime of motions, space-time measures which, in first approximation reduce to Lorentz transformations, do not diverge for $v=c$, and keep invariant light speed in vacuum for inertial observers under uniform relative motion, thus preserving causality. This set of space-time {\it measures} are given explicitly by eqs.(\ref{Leps-1}),(\ref{Leps-2}), or by  eqs.(\ref{ecsxt1}),(\ref{ecsyz1}) with $k=1$. In this regime of motions, called here the subluminal regime, $c_1=c$, $\gamma_1$ is given by eq.(\ref{gamma1}) and $\beta_1$ by eq.(\ref{ecsyz1}) with $u_1=v/c_1$. However, we see also that we can take other values of $k$ in eqs.(\ref{ecsxt1}),(\ref{ecsyz1}), for instance $k=2$, or even we can consider higher (positive) integer values with appropriate values of $\gamma_k$, $\beta_k$ and $c_k$, which correspond to what we call here superluminal regimes, being the first one that associated to the range of speeds $c<v\leq c_2$, where $\gamma_2$, $c_2$ are given by eqs.(\ref{gamma2}),(\ref{c_2}), respectively; the factor $\beta_2$ is calculated using eq.(\ref{ecsyz1}) with $u_2=v/c_2$.\\

For arbitrary $k\geq2$, that is, for all superluminal regimes, we have $u_k=v/c_k$, and $\gamma_k$, $c_k$ are given by eqs.(\ref{gammai}),(\ref{luz-i}), respectively, and $\beta_k$ is given by eq.(\ref{ecsyz1}). If we write $u$ without the subindex $k$, it means $u=v/c$ for any range of speeds we are dealing with, as for instance in the expressions for $\gamma_1$, $\gamma_2$, and $\gamma_k$. Superluminal regimes have speed intervals of type $c_{k-1}<v\leq c_k$, such that the top (and invariant) speed of the range approximately equals $c_k\simeq\epsilon^{-k+1}c$ -cf. eq.(\ref{luz-i}); therefore, $c_{k+1}/c_k\simeq\epsilon^{-1}$, thus $c_{k+1}\gg c_k$. At speed $c_k$, the top speed of any speeds range, the $\gamma_k$ factor takes the value $k\epsilon^{-1}$; we interpret here the states with speed $c_k$ as the speeds of the associated metrical null cones, which determine the description of any material body depending on its speed with respect to them. An interpretation for these special states associated to the signals propagating at speed $c_k$ was done, which behave as phenomena of one degree of freedom with respect to the direction of propagation and show a periodic behaviour.\\

Based on the results developed in present work, we have then generalized the ``local causality" postulate of space-time given by Hawking and Ellis, by asserting that the local causality holds for all motion regimes, the subluminal one and all superluminals, and the partition of motion regimes are given by a denumerable set of metrical null cones, each one characterized by some top and invariant speed $c_k$. We introduce a k-metric in eq.(\ref{k-metric}), where the natural number $k$ identifies the motion regime such that $k=1$ corresponds to the subluminal regime. We talk of k-timelike and k-null intervals -cf. eq.(\ref{k-interval}), as well as k-timelike and k-null paths which can be traced from/to a given event. Thus, we can connect event pairs through these paths and obtain a partial ordering in space-time, such that we can speak that some event $p$ causally or chronologically precedes other event $q$. With these tools in hand we have developed a causal structure of our space-time. Past and future regions of space-time causally or chronologically related to some given event $p$, were also considered for all motion regimes. Here we denote them by $J_k^{\pm}(p)$, $I_k^{\pm}(p)$, respectively, where the plus sign denotes future zones whilst the negative sign stands for the past ones. 

We also estimate a value for our constant $\epsilon^2\simeq3.7\times10^{-54}$, based on simple arguments which take into account the ideas of \citet{Gravitation} who consider the nature of vacuum fluctuations as fluctuations occurring at Planck levels of distances, and also arguing that they can be seen as an anomalous magnetic moment of electron; in this case we take into account the finest partition of space as Planck length.  

Finally, we can say that our space-time measures given by eqs.(\ref{ecsxt1}),(\ref{ecsyz1}) and the associated causal structure, enable us to think of making a discrete transition between the subluminal and the first superluminal regime as stipulated by eq.(\ref{cambio_u2}). To properly carry out it, one needs new dynamical measures compatible with our spacetime measures, indicating us how to carry out it, so work in this direction should be done. In a forthcoming work, the author will present dynamical measures compatible with present space-time measures, so a way to think of this kind of transition can be theoretically explored.

\section{Acknowledgementes}
The author thanks to J.C. Buitrago-Casas, MSc., (now at SSL, UC Berkeley) for making the digital version of present work figures.

\bibliography{art_BCM2014a.bib}

\end{document}